\documentclass[superscriptaddress,prb,showpacs,twocolumn]{revtex4}

\usepackage{amsmath}
\usepackage{amssymb}
\usepackage{epsfig}
\usepackage{graphicx}
\usepackage{wasysym}
\usepackage{multirow}

\newcommand \be{\begin{equation}}
\newcommand \ee{\end{equation}}
\newcommand \bes{\begin{equation*}} 
\newcommand \ees{\end{equation*}}
\newcommand \bea{\begin{eqnarray}}
\newcommand \eea{\end{eqnarray}}
\newcommand \beas{\begin{eqnarray*}} 
\newcommand \eeas{\end{eqnarray*}}
\newcommand \bfg{\begin{figure}}
\newcommand \efg{\end{figure}}
\newcommand \bfgs{\begin{figure*}} 
\newcommand \efgs{\end{figure*}}
\newcommand \bwt{\begin{widetext}}
\newcommand \ewt{\end{widetext}}

\newcommand \bra{\langle}
\newcommand \ket{\rangle}
\newcommand{\Fig}[1]{FIG.~\ref{#1}}
\newcommand \Equ[1]{(\ref{#1})}
\newcommand \Tab[1]{TABLE~\ref{#1}}
\newcommand \Sec[1]{Section~\ref{#1}}
\newcommand \SSec[1]{Subsection~\ref{#1}}
\newcommand \App[1]{Appendix~\ref{#1}}

\newcommand \vecS{{\bf S}}
\newcommand \vecT{{\bf T}}
\newcommand \vect{\vec{\tau}}

\newcommand \im{{i}}
\newcommand \dif{{\rm d}}
\newcommand \sgn{{\rm sgn}}

\newcommand \us{\uparrow}
\newcommand \ds{\downarrow}

\newcommand \etal{{\it et al.}}

\begin{document}

\title{Realization of the Exactly Solvable Kitaev Honeycomb Lattice Model in 
a Spin Rotation Invariant System}
\author{Fa Wang}
\affiliation{Department of Physics, Massachusetts Institute of Technology, 
Cambridge, MA 02139, USA}

\begin{abstract}
The exactly solvable Kitaev honeycomb lattice model is realized as the 
low energy effect Hamiltonian of 
a spin-1/2 model with spin rotation and time-reversal symmetry. 
The mapping to low energy effective Hamiltonian is exact, 
without truncation errors in traditional perturbation series expansions. 
This model consists of a honeycomb lattice of clusters 
of four spin-1/2 moments, and 
contains short-range interactions up to six-spin(or eight-spin) terms. 
The spin in the Kitaev model is represented not as these spin-1/2 moments,
but as pseudo-spin of the two-dimensional spin singlet sector of 
the four antiferromagnetically coupled spin-1/2 moments within each cluster.  
Spin correlations in the Kitaev model are mapped to 
dimer correlations or spin-chirality correlations in this model. 
This exact construction is quite general 
and can be used to make other interesting spin-1/2 models from 
spin rotation invariant Hamiltonians. 
We discuss two possible routes to generate the high order spin interactions 
from more natural couplings, which involves perturbative expansions 
thus breaks the exact mapping, although in a controlled manner.
\end{abstract}
\maketitle

\tableofcontents

\section{Introduction.}\label{sec:intro}
\begin{figure}
\includegraphics{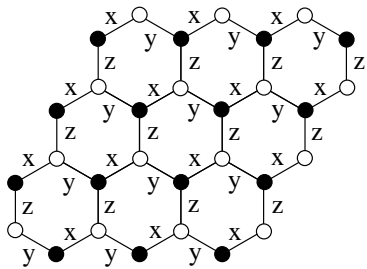}
\caption{The honeycomb lattice for the Kitaev model. 
Filled and open circles indicate two sublattices. 
$x,y,z$ label the links along three different directions
used in \Equ{equ:Kitaev}. 
}
\label{fig:kitaev}
\end{figure}

Kitaev's exactly solvable spin-1/2 honeycomb lattice model\cite{Kitaev} 
(noted as the Kitaev model hereafter)
has inspired great interest since its debut, 
due to its exact solvability, fractionalized excitations, 
and the potential to realize non-Abelian anyons. 
The model simply reads
\be
\begin{split}
H_{\rm Kitaev}=
 &
- \sum_{x-{\rm links}\ <jk>} J_x \tau^x_j\tau^x_{k}
- \sum_{y-{\rm links}\ <jk>} J_y \tau^y_j\tau^y_{k} 
\\ &
- \sum_{z-{\rm links}\ <jk>} J_z \tau^z_j\tau^z_{k}
\end{split}
\label{equ:Kitaev}
\ee
where $\tau^{x,y,z}$ are Pauli matrices, 
and $x,y,z$-links are defined in \Fig{fig:kitaev}. 
It was shown by Kitaev\cite{Kitaev} 
that this spin-1/2 model can be mapped to a model with 
one Majorana fermion per site coupled to Ising gauge fields on the links. 
And as the Ising gauge flux has no fluctuation, 
the model can be regarded as, under each gauge flux configuration, 
a free Majorana fermion problem. 
The ground state is achieved in the sector of zero gauge flux 
through each hexagon. 
The Majorana fermions in this sector have Dirac-like gapless dispersion 
resembling that of graphene, 
as long as $|J_x|$, $|J_y|$, and $|J_z|$ satisfy the triangular relation, 
sum of any two of them is greater than the third one\cite{Kitaev}. 
It was further proposed by Kitaev\cite{Kitaev} that 
opening of fermion gap by magnetic field 
can give the Ising vortices non-Abelian anyonic statistics, 
because the Ising vortex will carry a zero-energy Majorana mode, 
although magnetic field destroys the exact solvability. 

Great efforts have been invested to better understand 
the properties of the Kitaev model. 
For example, several groups have pointed out that 
the fractionalized Majorana fermion excitations 
may be understood from the more familiar Jordan-Wigner 
transformation of 1D spin systems\cite{XiangT, ChenHD}.
The analogy between the non-Abelian Ising vortices and 
vortices in $p+i p$ superconductors has been raised in serveral 
works\cite{LeeDH,YuY,YuYWangZQ,Kells}. 
Exact diagonalization has been used to study the Kitaev model on small 
lattices\cite{ChenHDProbe}. 
And perturbative expansion methods have been developed to study 
the gapped phases of the Kitaev-type models\cite{Schmidt}. 

Many generalizations of the Kitaev model have been derived as well. 
There have been several proposals 
to open the fermion gap for the non-Abelian phase 
without spoiling exact solvability\cite{LeeDH, YuYWangZQ}. 
And many generalizations to other(even 3D) lattices have been 
developed in the last few 
years\cite{Hong, SunCP, Hong2, Nussinov, Congjun, Ryu, Baskaran}. 
All these efforts 
have significantly enriched our knowledge of exactly solvable models 
and quantum phases of matter.

However, in the original Kitaev model and its later generalizations 
in the form of spin models, spin rotation symmetry 
is explicitly broken. 
This makes them harder to realize in solid state systems. 
There are many proposals to realized the Kitaev model in 
more controllable situations,
{e.g.} in cold atom optical lattices\cite{DuanLM,Micheli}, 
or in superconducting circuits\cite{YouJQ}.
But it is still desirable for theoretical curiosity and 
practical purposes to realize 
the Kitaev-type models in spin rotation invariant systems. 

In this paper we realize the Kitaev honeycomb lattice model as 
the low energy Hamiltonian for a spin rotation invariant system. 
The trick is {\em not} to use the physical spin as the spin in 
the Kitaev model, 
instead the spin-1/2 in Kitaev model is from some emergent two-fold degenerate 
low energy states in the elementary unit of physical system. 
This type of idea has been explored recently by Jackeli and Khaliullin
\cite{Khaliullin}, in which the spin-1/2 in the Kitaev model is 
the low energy Kramers doublet created by 
strong spin-orbit coupling of $t_{2g}$ orbitals. 
In the model presented below, the Hilbert space of spin-1/2 in 
the Kitaev model is actually the two dimensional spin singlet sector of 
four antiferromagnetically coupled spin-1/2 moments, 
and the role of spin-1/2 operators(Pauli matrices) in the Kitaev model is 
replaced by certain combinations of $\vecS_j\cdot \vecS_k$
[or the spin-chirality $\vecS_j\cdot(\vecS_k\times\vecS_{\ell})$] between 
the four spins. 

One major drawback of the model to be presented is that it contains 
high order spin interactions(involves up to six or eight spins), 
thus is still unnatural. 
However it opens the possibility to realize exotic (exactly solvable) models 
from spin-1/2 Hamiltonian with spin rotation invariant interactions. 
We will discuss two possible routes to reduce this artificialness through 
controlled perturbative expansions, by coupling to optical phonons or
by magnetic couplings between the elementary units.

The outline of this paper is as follows. In \Sec{sec:formulation} we will 
lay out the pseudo-spin-1/2 construction. In \Sec{sec:realize} the Kitaev 
model will be explicitly constructed using this formalism, and some properties
of this construction will be discussed. In \Sec{sec:perturb} we will discuss
two possible ways to generate the high order spin interactions involved 
in the construction of \Sec{sec:realize} by perturbative expansions. 
Conclusions and outlook will be summarized in \Sec{sec:conclusion}.

\section{Formulation of the Pseudo-spin-1/2 from Four-spin Cluster.}
\label{sec:formulation}
In this Section we will construct the pseudo-spin-1/2 from 
a cluster of four physical spins, and map the physical spin operators 
to pseudo-spin operators. The mapping constructed here will be used in later
Sections to construct the effective Kitaev model. 
In this Section we will work entirely within the four-spin cluster, 
all unspecified physical spin subscripts take values $1,\dots,4$. 

\begin{figure}
\includegraphics[scale=0.6]{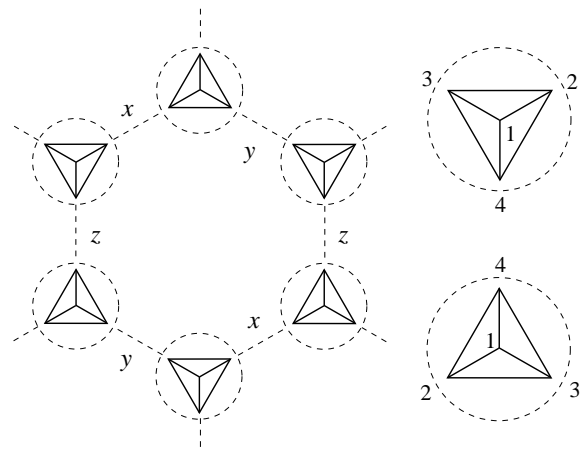}
\caption{Left: the physical spin lattice for the model \Equ{equ:Hexplicit}. 
The dash circles are honeycomb lattice sites, 
each of which is actually a cluster of four physical spins. 
The dash straight lines are honeycomb lattice bonds, 
with their type $x,y,z$ labeled. 
The interaction between clusters connected by $x,y,z$ bonds are the 
$J_{x,y,z}$ terms in \Equ{equ:Hexplicit} or \Equ{equ:Hexplicit2} respectively. 
Note this is not the 3-12 lattice used in Ref.~\cite{Hong,Schmidt}. 
Right: enlarged picture of the clusters with the four physical spins labeled 
as $1,\dots,4$. Thick solid bonds within one cluster have large 
antiferromagnetic Heisenberg coupling $J_{\rm cluster}$. 
}
\label{fig:Lattice}
\end{figure} 

Consider a cluster of four spin-1/2 moments(called physical spins hereafter), 
labeled by $\vecS_{1,\dots,4}$, antiferromagnetically coupled to each other
(see the right bottom part of \Fig{fig:Lattice}). 
The Hamiltonian within the cluster(up to a constant) is simply 
the Heisenberg antiferromagnetic(AFM) interactions, 
\be
H_{\rm cluster}=(J_{\rm cluster}/2)\left (
\vecS_1+\vecS_2+\vecS_3+\vecS_4
\right )^2
\label{equ:Hcluster}
\ee
The energy levels should be apparent from this form: 
one group of spin-2 quintets with energy $3J_{\rm cluster}$, 
three groups of spin-1 triplets with energy $J_{\rm cluster}$, 
and two spin singlets with energy zero. We will consider large positive
$J_{\rm cluster}$ limit. 
So only the singlet sector remains in low energy. 

The singlet sector is then 
treated as a pseudo-spin-1/2 Hilbert space. 
From now on we denote the pseudo-spin-1/2 operators as $\vecT=(1/2)\vect$, 
with $\vect$ the Pauli matrices. 
It is convenient to choose the following basis of the pseudo-spin
\be
\begin{split}
|\tau^z=\pm 1\ket= \ 
&
\frac{1}{\sqrt{6}}\Big (
 |\ds\ds\us\us\ket +\omega^{-\tau^z} |\ds\us\ds\us\ket + \omega^{\tau^z} |\ds\us\us\ds\ket
\\ \ &\quad
+|\us\us\ds\ds\ket +\omega^{-\tau^z} |\us\ds\us\ds\ket + \omega^{\tau^z} |\us\ds\ds\us\ket
\Big )
\end{split}
\label{equ:basis}
\ee
where $\omega=e^{2\pi\im/3}$ is the complex cubic root of unity, 
$|\ds\ds\us\us\ket$ and other states on the right-hand-side(RHS) are 
basis states of the four-spin system, in terms of $S^z$ quantum numbers 
of physical spins $1,\dots,4$ in sequential order. 
This pseudo-spin representation has been used by Harris {\etal} to study 
magnetic ordering in pyrochlore antiferromagnets\cite{Harris}. 

We now consider the effect of 
Heisenberg-type interactions $\vecS_j\cdot \vecS_k$ inside 
the physical singlet sector. 
Note that since any $\vecS_j\cdot \vecS_k$ within 
the cluster commutes with the cluster Hamiltonian $H_{\rm cluster}$ 
\Equ{equ:Hcluster}, 
their action do not mix physical spin singlet states with states of other 
total physical spin. This property is also true for 
the spin-chirality operator used later. 
So the pseudo-spin Hamiltonian constructed below will be {\em exact} 
low energy Hamiltonian, 
without truncation errors in typical perturbation series expansions.

It is simpler to consider the permutation operators
 $P_{jk}\equiv 2\vecS_{j}\cdot \vecS_{k}+1/2$, which just exchange 
the states of the two physical spin-1/2 moments $j$ and $k$ ($j\neq k$). 
As an example we consider the action of $P_{34}$, 
\bes
\begin{split}
P_{34}|\tau^z=-1\ket
=\ 
&
 \frac{1}{\sqrt{6}}
\Big (
 |\ds\ds\us\us\ket +\omega |\ds\us\us\ds\ket + \omega^2 |\ds\us\ds\us\ket
\\ \ &\quad
+|\us\us\ds\ds\ket +\omega |\us\ds\ds\us\ket + \omega^2 |\us\ds\us\ds\ket
\Big )
\\
=\ 
 &
|\tau^z=+1\ket
\end{split}
\ees
and similarly $P_{34}|\tau^z=-1\ket=|\tau^z=+1\ket$. 
Therefore $P_{34}$ is just $\tau^x$ in the physical singlet sector. 
A complete list of all permutation operators 
is given in \Tab{table:correspondence}.
We can choose the following representation of 
$\tau^x$ and $\tau^y$, 
\be
\begin{split}
 &
\tau^x=P_{12}=2\vecS_{1}\cdot\vecS_2+1/2
\\ &
 \tau^y=(P_{13}-P_{14})/\sqrt{3} = (2/\sqrt{3})\vecS_{1}\cdot(\vecS_{3}-\vecS_{4})
\end{split}
\label{equ:tauxy}
\ee
Many other representations are possible as well, 
because several physical spin interactions may correspond to the same
pseudo-spin interaction in the physical singlet sector, 
and we will take advantage of this later.

For $\tau^z$ we can use $\tau^z=-\im \tau^x\tau^y$, where $\im$ is 
the imaginary unit,  
\be
\tau^z=-\im(2/\sqrt{3})(2\vecS_{1}\cdot\vecS_2+1/2)\vecS_{1}\cdot(\vecS_{3}-\vecS_{4})
\label{equ:tauz1}
\ee
However there is another simpler representation of $\tau^z$, 
by the spin-chirality operator $\chi_{jk\ell}=\vecS_j\cdot(\vecS_k\times\vecS_{\ell})$. 
Explicit calculation shows that the effect of 
$\vecS_2\cdot(\vecS_3\times\vecS_4)$ is $-(\sqrt{3}/4)\tau^z$ in the 
physical singlet sector. 
This can also be proved by using the commutation relation 
$[\vecS_{2}\cdot\vecS_{3},\vecS_{2}\cdot\vecS_{4}]=
\im \vecS_{2}\cdot(\vecS_{3}\times\vecS_{4})$. 
A complete list of all chirality operators 
is given in \Tab{table:correspondence}.
Therefore we can choose another representation of $\tau^z$,
\be
\tau^z=-\chi_{234}/(\sqrt{3}/4)=-(4/\sqrt{3})\vecS_2\cdot(\vecS_3\times\vecS_4)
\label{equ:tauz2}
\ee
The above representations of $\tau^{x,y,z}$ are all 
invariant under global spin rotation of the physical spins.

\begin{table}[t]
\begin{tabular}{|l|l|}
\hline
physical spin & pseudo-spin \\
\hline\hline
$P_{12}$,\ {\rm and}\ $P_{34}$ & $\tau^x$ \\
\hline
$P_{13}$,\ {\rm and}\ $P_{24}$ & $-(1/2)\tau^x+(\sqrt{3}/2)\tau^y$ \\
\hline
$P_{14}$,\ {\rm and}\ $P_{23}$ & $-(1/2)\tau^x-(\sqrt{3}/2)\tau^y$ \\
\hline
$-\chi_{234}$,\ $\chi_{341}$,\ $-\chi_{412}$,\ {\rm and}\ $\chi_{123}$ & $(\sqrt{3}/4)\tau^z$\\
\hline
\end{tabular}
\caption{
Correspondence between physical spin operators and pseudo-spin operators 
in the physical spin singlet sector of the four antiferromagnetically coupled 
physical spins. 
$P_{jk}=2\vecS_j\cdot \vecS_k+1/2$ are permutation operators, 
$\chi_{jk\ell}=\vecS_j\cdot(\vecS_k\times\vecS_{\ell})$ are 
spin-chirality operators. 
Note that several 
physical spin operators may correspond to the same pseudo-spin operator. 
}
\label{table:correspondence}
\end{table}

With the machinery of equations \Equ{equ:tauxy}, \Equ{equ:tauz1}, and 
\Equ{equ:tauz2}, it will be straightforward to 
construct various pseudo-spin-1/2 Hamiltonians on various lattices, 
of the Kitaev variety and beyond,
as the exact low energy effective Hamiltonian of certain spin-1/2 models
with spin-rotation symmetry.   
In these constructions a pseudo-spin lattice site actually represents
a cluster of four spin-1/2 moments. 

\section{Realization of the Kitaev Model.}\label{sec:realize}
In this Section we will use directly the results of the previous Section
to write down a Hamiltonian whose low energy sector is described 
by the Kitaev model. The Hamiltonian will be constructed on the 
physical spin lattice illustrated in \Fig{fig:Lattice}. 
In this Section we will use $j,k$ to label four-spin clusters 
(pseudo-spin-1/2 sites), the physical spins in cluster $j$ are labeled as
$\vecS_{j1},\dots,\vecS_{j4}$. 

Apply the mappings developed in \Sec{sec:formulation}, we have the 
desired Hamiltonian in short notation, 
\be
\begin{split}
H=
 & 
\sum_{{\rm cluster}}H_{{\rm cluster}}
- \sum_{x-{\rm links}\ <jk>} J_x \tau^x_j\tau^x_{k}
\\ &
- \sum_{y-{\rm links}\ <jk>} J_y \tau^y_j\tau^y_{k}
- \sum_{z-{\rm links}\ <jk>} J_z \tau^z_j\tau^z_{k}
\end{split}
\label{equ:H}
\ee
where $j,k$ label the honeycomb lattice sites thus the four-spin clusters, 
$H_{\rm cluster}$ is given by \Equ{equ:Hcluster}, 
$\tau^{x,y,z}$ should be replaced by the corresponding physical spin 
operators in \Equ{equ:tauxy} and \Equ{equ:tauz1} or \Equ{equ:tauz2}, 
or some other equivalent representations of personal preference. 

Plug in the expressions \Equ{equ:tauxy} and \Equ{equ:tauz2} into \Equ{equ:H}, 
the Hamiltonian reads explicitly as
\bwt
\be
\begin{split}
H=\ 
&
\sum_{j}
(J_{\rm cluster}/2)(\vecS_{j1}+\vecS_{j2}+\vecS_{j3}+\vecS_{j4})^2
-\sum_{z-{\rm links}\ <jk>}J_z\,
(16/9)[\vecS_{j2}\cdot(\vecS_{j3}\times\vecS_{j4})]
[\vecS_{k2}\cdot(\vecS_{k3}\times\vecS_{k4})]
\\ \ &
-\sum_{x-{\rm links}\ <jk>}J_x\,
 (2\vecS_{j1}\cdot\vecS_{j2}+1/2)
 (2\vecS_{k1}\cdot\vecS_{k2}+1/2)
-\sum_{y-{\rm links}\ <jk>}J_y\,
(4/3)[\vecS_{j1}\cdot(\vecS_{j3}-\vecS_{j4})]
[\vecS_{k1}\cdot(\vecS_{k3}-\vecS_{k4})]
\end{split}
\label{equ:Hexplicit}
\ee
\ewt

While by the represenation \Equ{equ:tauxy} and \Equ{equ:tauz1}, 
the Hamiltonian becomes
\bwt
\be
\begin{split}
H=\ 
&
\sum_{j}
(J_{\rm cluster}/2)(\vecS_{j1}+\vecS_{j2}+\vecS_{j3}+\vecS_{j4})^2
\\ \ &
-\sum_{x-{\rm links}\ <jk>}J_x\,
 (2\vecS_{j1}\cdot\vecS_{j2}+1/2)
 (2\vecS_{k1}\cdot\vecS_{k2}+1/2)
-\sum_{y-{\rm links}\ <jk>}J_y\,
(4/3)[\vecS_{j1}\cdot(\vecS_{j3}-\vecS_{j4})]
[\vecS_{k1}\cdot(\vecS_{k3}-\vecS_{k4})] 
\\ \ &
-\sum_{z-{\rm links}\ <jk>}J_z\,
(-4/3) (2\vecS_{j3}\cdot\vecS_{j4}+1/2)[\vecS_{j1}\cdot(\vecS_{j3}-\vecS_{j4})]
 (2\vecS_{k3}\cdot\vecS_{k4}+1/2)[\vecS_{k1}\cdot(\vecS_{k3}-\vecS_{k4})] 
\end{split}
\label{equ:Hexplicit2}
\ee
\ewt

This model, in terms of physical spins $\vecS$, has full spin rotation symmetry 
and time-reversal symmetry. 
A pseudo-magnetic field term $\sum_{j}\vec{h}\cdot\vect_j$ term can also be 
included under this mapping, however the resulting Kitaev model 
with magnetic field is not exactly solvable. 
It is quite curious that such a formidably looking Hamiltonian
 \Equ{equ:Hexplicit},
with biquadratic and six-spin(or eight-spin) terms, 
has an exactly solvable low energy sector. 

We emphasize that because 
the first intra-cluster term $\sum_{{\rm cluster}}H_{{\rm cluster}}$ 
commutes with the latter Kitaev terms independent of the representation used, 
the Kitaev model is realized as the {\em exact} low energy Hamiltonian of 
this model without truncation errors of perturbation theories, 
namely no $(|J_{x,y,z}|/J_{\rm cluster})^2$ or higher order terms will be
generated under the projection to low energy cluster singlet space. 
This is unlike, for example, the $t/U$ expansion of the half-filled Hubbard 
model\cite{Oles,MacDonald},
where at lowest $t^2/U$ order the effective Hamiltonian is 
the Heisenberg model, but higher order terms ($t^4/U^3$ etc.) 
should in principle still 
be included in the low energy effective Hamiltonian for any finite $t/U$. 
Similar comparison can be made to the perturbative expansion studies of 
the Kitaev-type models by Vidal {\etal}\cite{Schmidt}, 
where the low energy effective Hamiltonians were obtained 
in certian anisotropic (strong bond/triangle) limits. 
Although the spirit of this work, namely projection to low energy sector, 
is the same as all previous perturbative approaches to effective Hamiltonians. 

Note that the original Kitaev model \Equ{equ:Kitaev} has 
three-fold rotation symmetry around a honeycomb lattice site, combined with 
a three-fold rotation in pseudo-spin space
(cyclic permutation of $\tau^x$, $\tau^y$, $\tau^z$).
This is not apparent in our model \Equ{equ:Hexplicit} 
in terms of physical spins, under the current representation of $\tau^{x,y,z}$. 
We can remedy this by using a different set of pseudo-spin Pauli matrices 
$\tau'^{x,y,z}$ in \Equ{equ:H},  
\bes
\begin{split}
&
\tau'^x=\sqrt{1/3}\tau^z+\sqrt{2/3}\tau^x,\quad
\\ &
\tau'^y=\sqrt{1/3}\tau^z-\sqrt{1/6}\tau^x+\sqrt{1/2}\tau^y,
\\ &
\tau'^z=\sqrt{1/3}\tau^z-\sqrt{1/6}\tau^x-\sqrt{1/2}\tau^y
\end{split}
\ees
With proper representation choice, they have a symmetric form in
terms of physical spins, 
\be
\begin{split}
&
\tau'^x=-(4/3)\vecS_2\cdot(\vecS_3\times\vecS_4)+\sqrt{2/3}(2\vecS_1\cdot\vecS_2+1/2)
\\ &
\tau'^y=-(4/3)\vecS_3\cdot(\vecS_4\times\vecS_2)+\sqrt{2/3}(2\vecS_1\cdot\vecS_3+1/2)
\\ &
\tau'^z=-(4/3)\vecS_4\cdot(\vecS_2\times\vecS_3)+\sqrt{2/3}(2\vecS_1\cdot\vecS_4+1/2)
\end{split}
\ee
So the symmetry mentioned above can be realized by a three-fold rotation of 
the honeycomb lattice, with a cyclic permutation of $\vecS_2$, $\vecS_3$ and 
$\vecS_4$ in each cluster. This is in fact the three-fold rotation symmetry of 
the physical spin lattice illustrated in \Fig{fig:Lattice}. 
However this more symmetric 
representation will not be used in later part of this paper. 

Another note to take is that it is not necessary to have such 
a highly symmetric cluster Hamiltonian \Equ{equ:Hcluster}. 
The mappings to pseudo-spin-1/2 should work 
as long as the ground states of the cluster Hamiltonian are 
the two-fold degenerate singlets. 
One generalization, which conforms 
the symmetry of the lattice in \Fig{fig:Lattice}, is to have
\be
H_{\rm cluster}=(J_{\rm cluster}/2)(r\cdot\vecS_1+\vecS_2+\vecS_3+\vecS_4)^2
\ee
with $J_{\rm cluster}>0$ and $0<r<3$. 
However this is not convenient for later discussions and will not be used. 

We briefly describe some of the properties of \Equ{equ:Hexplicit}. 
Its low energy states are entirely in the space that 
each of the clusters is a physical spin singlet 
(called cluster singlet subspace hereafter).
Therefore physical spin correlations are strictly confined within 
each cluster. 
The excitations carrying physical spin are gapped, 
and their dynamics are `trivial' in the sense that they do not 
move from one cluster to another. 
But there are non-trivial low energy physical spin singlet excitations, 
described by the pseudo-spins defined above. 
The correlations of the pseudo-spins can be mapped to 
correlations of their corresponding physical spin observables 
(the inverse mappings are not unique, c.f. \Tab{table:correspondence}). 
For example $\tau^{x,y}$ correlations become certain dimer-dimer correlations, 
$\tau^{z}$ correlation becomes chirality-chirality correlation, 
or four-dimer correlation. 
It will be interesting to see the corresponding picture of 
the exotic excitations in the Kitaev model, 
{e.g.} the Majorana fermion and the Ising vortex. 
However this will be deferred to future studies. 

It is tempting to call this as an exactly solved spin liquid with spin gap 
($\sim J_{\rm cluster}$), 
an extremely short-range resonating valence bond(RVB) state, 
from a model with spin rotation and time reversal symmetry. 
However it should be noted that the unit cell of this model contains 
an even number of spin-1/2 moments (so does the original Kitaev model) 
which does not satisfy the stringent definition of spin liquid 
requiring odd number of electrons per unit cell. 
Several parent Hamiltonians of spin liquids have already been constructed. 
See for example, Ref.~\cite{ChayesKivelson, Batista, RamanSondhi, Thomale}.

\section{Generate the High Order Physical Spin Interactions 
by Perturbative Expansion.}\label{sec:perturb}

One major drawback of the present construction is 
that it involves high order interactions of  
physical spins[see \Equ{equ:Hexplicit} and \Equ{equ:Hexplicit2}], 
thus is `unnatural'. 
In this Section we will make compromises between exact solvability and 
naturalness. 
We consider two clusters $j$ and $k$ and try to generate the $J_{x,y,z}$ 
interactions in \Equ{equ:H} from perturbation series expansion 
of more natural(lower order) physical spin interactions. 
Two different approaches for this purpose will be laid out in 
the following two Subsections. 
In \SSec{ssec:phonon} we will consider the two clusters as two 
tetrahedra, and couple the spin system to certain optical phonons, 
further coupling between the phonon modes of the two clusters 
can generate at lowest order the desired high order spin interactions. 
In \SSec{ssec:magnetic} we will introduce certain magnetic, 
{e.g.} Heisenberg-type, interactions between physical spins of 
different clusters, at lowest order(second order) of perturbation theory 
the desired high order spin interactions can be achieved.
These approaches involve truncation errors in the perturbation series, 
thus the mapping to low energy effect Hamiltonian will no longer be exact. 
However the error introduced may be controlled by 
small expansion parameters. 
In this Section we denote the physical spins on cluster $j$($k$) 
as $j1,\dots,j4$ ($k1,\dots,k4$), 
and denote pseudo-spins on cluster $j$($k$) as $\vect_j$($\vect_k$).

\subsection{Generate the High Order Terms by Coupling to Optical Phonon.}
\label{ssec:phonon}
In this Subsection we regard each four-spin cluster as a tetrahedron, 
and consider possible optical phonon modes(distortions) and their 
couplings to the spin system. 
The basic idea is that the intra-cluster Heisenberg coupling $J_{\rm cluster}$ 
can linearly depend on the distance between physical spins. 
Therefore certain distortions of the tetrahedron 
couple to certain linear combinations of $\vecS_{\ell}\cdot \vecS_m$. 
Integrating out phonon modes will then generate high order spin 
interactions. 
This idea has been extensively studied and 
applied to several magnetic 
materials\cite{Tchernyshyov, Mila, Penc, Mila2, Tchernyshyov2, Bergman, FW}. 
More details can be found in a recent review by 
Tchernyshyov and Chern\cite{TchernyshyovReview}. 
And we will frequently use their notations. 
In this Subsection we will use the representation 
\Equ{equ:tauz1} for $\tau^z$. 

\begin{figure}
\includegraphics[scale=0.5]{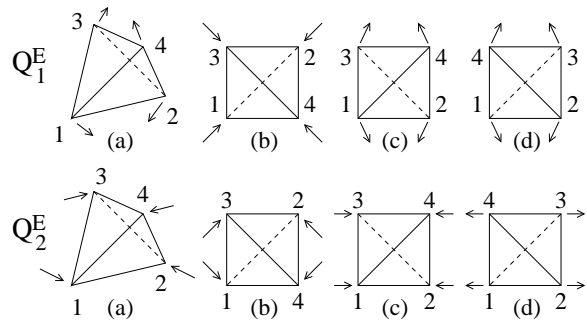}
\caption{Illustration of the tetragonal to orthorhombic 
$Q^E_1$(top) and $Q^E_2$(bottom) distortion modes. 
(a) Perspective view of the tetrahedron. $1,\dots,4$ label the 
spins. Arrows indicate the motion of each spin under the distortion mode. 
(b) Top view of (a). (c)(d) Side view of (a).
}
\label{fig:QE}
\end{figure}

Consider first a single tetrahedron with four spins $1,\dots,4$. 
The general distortions of this tetrahedron can be classified by their symmetry
(see for example Ref.~\cite{TchernyshyovReview}). 
Only two tetragonal to orthorhombic distortion modes, 
$Q^E_1$ and $Q^E_2$ (illustrated in \Fig{fig:QE}), 
couple to the pseudo-spins defined in \Sec{sec:formulation}. 
A complete analysis of all modes is given in \App{app:phonon}. 
The coupling is of the form
\bes
J'(Q^E_1 f^E_1+Q^E_2 f^E_2)
\ees
where $J'$ is the derivative of Heisenberg coupling $J_{\rm cluster}$ 
between two spins $\ell$ and $m$ 
with respect to their distance $r_{\ell m}$,
 $J'= \dif J_{\rm cluster}/\dif r_{\ell m}$;
$Q^E_{1,2}$ are the generalized coordinates of these two modes; 
and the functions $f^E_{1,2}$ are 
\bes
\begin{split}
f^E_2\ & = (1/2)(
 \vecS_2\cdot\vecS_4+\vecS_1\cdot\vecS_3
-\vecS_1\cdot\vecS_4-\vecS_2\cdot\vecS_3
),
\\
f^E_1\ & = \sqrt{1/12}(
 \vecS_1\cdot\vecS_4+\vecS_2\cdot\vecS_3+\vecS_2\cdot\vecS_4
+\vecS_1\cdot\vecS_3
\\ & \phantom{=\sqrt{1/12}(}
-2\vecS_1\cdot\vecS_2-2\vecS_3\cdot\vecS_4
).
\end{split}
\ees
According to \Tab{table:correspondence} we have 
$f^E_1=-(\sqrt{3}/2)\tau^x$
and 
$f^E_2=(\sqrt{3}/2)\tau^y$. Then the coupling becomes 
\be
(\sqrt{3}/2)J'(-Q^E_1\, \tau^x+Q^E_2\, \tau^y)
\label{equ:HSLcluster}
\ee

The spin-lattice(SL) Hamiltonian on a single cluster $j$ is 
[equation (1.8) in Ref.~\cite{TchernyshyovReview}],
\be
\begin{split}
H_{{\rm cluster\ }j,{\rm\ SL}}=
 & 
H_{{\rm cluster}\ j}
+\frac{k}{2}(Q^E_{1j})^2+\frac{k}{2}(Q^E_{2j})^2
\\ &
-\frac{\sqrt{3}}{2}J'(Q^E_{1j} \tau^x_j-Q^E_{2j}\tau^y_j),
\end{split}
\label{equ:HclusterSL}
\ee
where $k>0$ is the elastic constant for these phonon modes, 
$J'$ is the spin-lattice coupling constant, 
$Q^E_{1j}$ and $Q^E_{2j}$ are the generalized coordinates of the 
$Q^E_1$ and $Q^E_2$ distortion modes of cluster $j$, 
$H_{{\rm cluster}\ j}$ is \Equ{equ:Hcluster}. 
As already noted in Ref.~\cite{TchernyshyovReview}, 
this model does not really break the pseudo-spin rotation symmetry 
of a single cluster. 

Now we put two clusters $j$ and $k$ together, 
and include a perturbation $\lambda\, H_{\rm perturbation}$ to 
the optical phonon Hamiltonian, 
\bes
\begin{split}
H_{jk,{\rm SL}}=
&
H_{{\rm cluster\ }j,{\rm\ SL}}
+H_{{\rm cluster\ }k,{\rm\ SL}}
\\ &
+\lambda\, H_{\rm perturbation}[Q^E_{1j},Q^E_{2j},Q^E_{1k},Q^E_{2k}]
\end{split}
\ees
where $\lambda$ (in fact $\lambda/k$) is the expansion parameter.

Consider the perturbation
 $H_{\rm perturbation}=Q^E_{1j}\cdot Q^E_{1k}$, 
which means a coupling between the $Q^E_1$ distortion modes of 
the two tetrahedra. 
Integrate out the optical phonons, 
at lowest non-trivial order, 
it produces a term $(3\,J'^2\,\lambda)/(4\, k^2)\,\tau^x_j\cdot \tau^x_k$. 
This can be seen by minimizing separately the two cluster Hamiltonians 
with respect to $Q^E_1$, which gives 
$Q^E_1=(\sqrt{3}\,J')/(2\,k)\tau^x$, 
then plug this into the perturbation term.
Thus we have produced the $J_x$ term in the Kitaev model with
 $J_x=-(3\,J'^2\,\lambda)/(4\, k^2)$. 

Similarly the perturbation
 $H_{\rm perturbation}=Q^E_{2j}\cdot Q^E_{2k}$
will generate 
 $(3\,J'^2\,\lambda)/(4\, k^2)\,\tau^y_j\cdot \tau^y_k$ 
at lowest non-trivial order. 
So we can make $J_y=-(3\,J'^2\,\lambda)/(4\, k^2)$. 

The $\tau^z_j\cdot \tau^z_k$ coupling is more difficult to get.  
We treat it as $-\tau^x_j\tau^y_j\cdot \tau^x_k\tau^y_k$. 
By the above reasoning, we need an anharmonic coupling 
 $H_{\rm perturbation}=Q^E_{1j}Q^E_{2j}\cdot Q^E_{1k}Q^E_{2k}$. 
It will produce at lowest non-trivial order
 $(9\,J'^4\,\lambda)/(16\,k^4)\,\tau^x_j\tau^y_j\cdot \tau^x_k\tau^y_k$. 
Thus we have $J_z=(9\,J'^4\,\lambda)/(16\,k^4)$. 

Finally we have made up a spin-lattice model $H_{\rm SL}$, 
which involves only $\vecS_{\ell}\cdot\vecS_m$ interaction for physical spins, 
\bes
\begin{split}
H_{\rm SL}=
&
\sum_{\rm cluster}H_{{\rm cluster},\ {\rm SL}}
+\sum_{x-{\rm links}\ <jk>}\lambda_x\, Q^E_{1j}\cdot Q^E_{1k}
\\ &
+\sum_{y-{\rm links}\ <jk>}\lambda_y\, Q^E_{2j}\cdot Q^E_{2k} 
\\ &
+\sum_{z-{\rm links}\ <jk>}\lambda_z\, Q^E_{1j}Q^E_{2j}\cdot Q^E_{1k}Q^E_{2k}
\end{split}
\ees
where $Q^E_{1j}$ is the generalized coordinate for the $Q^E_1$ mode 
on cluster $j$, and $Q^E_{1k}$, $Q^E_{2j}$, $Q^E_{2k}$ are similarly defined; 
 $\lambda_{x,y}=-(4 J_{x,y} k^2)/(3J'^2)$ and
 $\lambda_z=(16J_z k^4)/(9J'^4)$; 
the single cluster spin-lattice Hamiltonian $H_{{\rm cluster},\ {\rm SL}}$ is 
\Equ{equ:HclusterSL}.

Collect the results above we have the spin-lattice Hamiltonian $H_{\rm SL}$ 
explicitly written as, 
\bwt
\be
\begin{split}
H_{\rm SL}
=
\ &
\sum_{{\rm cluster}\ j}\Big [
(J_{\rm cluster}/2)(
\vecS_{j1}+\vecS_{j2}+\vecS_{j3}+\vecS_{j4})^2
+
\frac{k}{2}(Q^E_{1j})^2+\frac{k}{2}(Q^E_{2j})^2
\\ & \phantom{ \sum_{{\rm cluster}\ j}\Big [ }
+ J'\Big ( 
Q^E_{1j}\frac{
 \vecS_{j1}\cdot\vecS_{j4}+\vecS_{j2}\cdot\vecS_{j3}+\vecS_{j2}\cdot\vecS_{j4}
+\vecS_{j1}\cdot\vecS_{j3}
-2\vecS_{j1}\cdot\vecS_{j2}-2\vecS_{j3}\cdot\vecS_{j4}
}{\sqrt{12}}
\\ & \phantom{ \sum_{{\rm cluster}\ j}\Big [ + J'\Big ( }
+Q^E_{2j}\frac{\vecS_{j2}\cdot\vecS_{j4}+\vecS_{j1}\cdot\vecS_{j3}
-\vecS_{j1}\cdot\vecS_{j4}-\vecS_{j2}\cdot\vecS_{j3}
}{2}
\Big )
\Big ]
\\ &
-\sum_{x-{\rm links}\ <jk>}\frac{4 J_x k^2}{3J'^2}\, Q^E_{1j}\cdot Q^E_{1k}
-\sum_{y-{\rm links}\ <jk>}\frac{4 J_y k^2}{3J'^2}\, Q^E_{2j}\cdot Q^E_{2k} 
+\sum_{z-{\rm links}\ <jk>}\frac{16 J_z k^4}{9J'^4}\, Q^E_{1j}Q^E_{2j}\cdot Q^E_{1k}Q^E_{2k}
\end{split}
\label{equ:HSLexplicit}
\ee
\ewt
The single cluster spin-lattice Hamiltonian 
[first three lines in \Equ{equ:HSLexplicit}] is quite natural. 
However we need some harmonic(on $x$- and $y$-links of honeycomb lattice) 
and anharmonic coupling (on $z$-links)
between optical phonon modes of neighboring tetrahedra. 
And these coupling constants $\lambda_{x,y,z}$ need to be tuned to produce 
$J_{x,y,z}$ of the Kitaev model. 
This is still not easy to implement in solid state systems. 
At lowest non-trivial order of perturbative expansion, 
we do get our model \Equ{equ:Hexplicit2}. 
Higher order terms in expansion destroy the exact solvability,  
but may be controlled by the small parameters $\lambda_{x,y,z}/k$. 

\subsection{Generate the High Order Terms by Magnetic Interactions
 between Clusters.}\label{ssec:magnetic}
In this Subsection we consider more conventional perturbations, 
magnetic interactions between the clusters, {e.g.} the Heisenberg coupling 
$\vecS_j\cdot\vecS_k$ with $j$ and $k$ belong to different tetrahedra. 
This has the advantage over the previous phonon approach for not introducing 
additional degrees of freedom. 
But it also has a significant disadvantage: 
the perturbation does not commute with the cluster Heisenberg Hamiltonian
\Equ{equ:Hcluster}, so the cluster singlet subspace will be mixed 
with other total spin states. 
In this Subsection we will use the spin-chirality representation 
\Equ{equ:tauz2} for $\tau^z$. 

Again consider two clusters $j$ and $k$. 
For simplicity of notations define a projection operator 
 $\mathcal{P}_{jk}=\mathcal{P}_{j}\mathcal{P}_{k}$, 
where $\mathcal{P}_{j,k}$ is projection into the singlet subspace of 
cluster $j$ and $k$, respectively, 
$
\mathcal{P}_{j,k}=\sum_{s=\pm 1}
|\tau^z_{j,k}=s\ket\bra\tau^z_{j,k}=s|
$.
For a given perturbation $\lambda\,H_{\rm perturbation}$ 
with small parameter $\lambda$ 
(in factor $\lambda/J_{\rm cluster}$ is the expansion parameter), 
lowest two orders of the perturbation series are
\be
\begin{split}
& 
\lambda\,\mathcal{P}_{jk}H_{\rm perturbation}\mathcal{P}_{jk}
+\lambda^2\,
\mathcal{P}_{jk}H_{\rm perturbation}(1-\mathcal{P}_{jk})
\\ &
\times
[0-H_{{\rm cluster}\ j}-H_{{\rm cluster}\ k}]^{-1}
(1-\mathcal{P}_{jk})H_{\rm perturbation}\mathcal{P}_{jk}
\end{split}
\label{equ:PerturbationSeries}
\ee
With proper choice of $\lambda$ and $H_{\rm perturbation}$ 
we can generate the desired $J_{x,y,z}$ terms in \Equ{equ:Hexplicit} 
from the first and second order of perturbations. 

The calculation can be dramatically simplified by the following fact 
that any physical spin-1/2 operator $S^{x,y,z}_{\ell}$ converts
the cluster spin singlet
states $|\tau^z=\pm 1\ket$ into spin-1 states of the cluster. 
This can be checked by explicit calculations and will not be proved here. 
For all the perturbations to be considered later, 
the above mentioned fact can be exploited to replace 
the factor
$[0-H_{{\rm cluster}\ j}-H_{{\rm cluster}\ k}]^{-1}$
in the second order perturbation 
to a $c$-number
$(-2J_{\rm cluster})^{-1}$. 

The detailed calculations are given in \App{app:derive}. 
We will only list the results here.

The perturbation on $x$-links is given by
\bes
\begin{split}
\lambda_x\, H_{{\rm perturbation},\ x}
=\ 
&
\lambda_x
[\vecS_{j1}\cdot\vecS_{k1}+ \sgn(J_x)\cdot(\vecS_{j2}\cdot\vecS_{k2})]
\\ &
-J_x(\vecS_{j1}\cdot\vecS_{j2}+\vecS_{k1}\cdot\vecS_{k2}).
\end{split}
\label{equ:xPerturbation}
\ees
where 
$\lambda_x=\sqrt{12 |J_x|\cdot J_{\rm cluster}}$, 
$\sgn(J_x)=\pm 1$ is the sign of $J_x$.

The perturbation on $y$-links is
\bes
\begin{split}
&
\lambda_y\, H_{{\rm perturbation},\ y}
\\
=
& 
\lambda_y
[\vecS_{j1}\cdot\vecS_{k1}+ {\rm sgn}(J_y)\cdot
(\vecS_{j3}-\vecS_{j4})\cdot(\vecS_{k3}-\vecS_{k4})]
\\ & 
-|J_y| (\vecS_{j3}\cdot \vecS_{j4}+\vecS_{k3}\cdot \vecS_{k4})
\end{split}
\label{equ:yPerturbation}
\ees
with $\lambda_y=\sqrt{4 |J_y|\cdot J_{\rm cluster}}$. 

The perturbation on $z$-links is
\bes
\begin{split}
& 
\lambda_z\, H_{{\rm perturbation},\ z}
\\
=\ 
& 
\lambda_z
[\vecS_{j2}\cdot(\vecS_{k3}\times\vecS_{k4})
+\sgn(J_z)\cdot\vecS_{k2}\cdot(\vecS_{j3}\times\vecS_{j4})]
\\ & 
-|J_z| (\vecS_{j3}\cdot \vecS_{j4}+\vecS_{k3}\cdot \vecS_{k4}).
\end{split}
\label{equ:zPerturbation}
\ees
with $\lambda_z=4\sqrt{|J_z|\cdot J_{\rm cluster}}$. 

The entire Hamiltonian $H_{\rm magnetic}$ reads explicitly as,
\bwt
\be
\begin{split}
& H_{\rm magnetic} = 
\sum_{{\rm cluster}\ j}
(J_{\rm cluster}/2)(
\vecS_{j1}+\vecS_{j2}+\vecS_{j3}+\vecS_{j4})^2
\\ &\quad
+\sum_{x-{\rm links}\ <jk>}
\big \{
\sqrt{12 |J_x|\cdot J_{\rm cluster}}
\big [
 \vecS_{j1}\cdot\vecS_{k1}+ \sgn(J_x)\cdot(\vecS_{j2}\cdot\vecS_{k2})
\big ]
-J_x(\vecS_{j1}\cdot\vecS_{j2}+\vecS_{k1}\cdot\vecS_{k2})
\big \}
\\ &\quad
+\sum_{y-{\rm links}\ <jk>}
\big \{
\sqrt{4 |J_y|\cdot J_{\rm cluster}}
\big [
 \vecS_{j1}\cdot(\vecS_{k3}-\vecS_{k4})+ \sgn(J_y)
 \vecS_{k1}\cdot(\vecS_{j3}-\vecS_{j4})
\big ]
 -|J_y| (\vecS_{j3}\cdot \vecS_{j4}+\vecS_{k3}\cdot \vecS_{k4})
\big \} 
\\ &\quad
+\sum_{z-{\rm links}\ <jk>}
\big \{
 4\sqrt{|J_z|\cdot J_{\rm cluster}}
\big [
 \vecS_{j2}\cdot(\vecS_{k3}\times\vecS_{k4})+\sgn(J_z)
 \vecS_{k2}\cdot(\vecS_{j3}\times\vecS_{j4})
\big ]
 -|J_z| (\vecS_{j3}\cdot \vecS_{j4}+\vecS_{k3}\cdot \vecS_{k4})
\big \}.
\end{split}
\label{equ:Hmagnetic}
\ee
\ewt
 
In \Equ{equ:Hmagnetic}, 
we have been able to reduce the four spin interactions in
 \Equ{equ:Hexplicit} to inter-cluster Heisenberg interactions, 
and the six-spin interactions in \Equ{equ:Hexplicit} to 
inter-cluster spin-chirality interactions. 
The inter-cluster Heisenberg couplings in $H_{{\rm perturbation}\ x,y}$ may be 
easier to arrange. 
The inter-cluster spin-chirality coupling in
 $H_{{\rm perturbation}\ z}$ explicitly breaks time reversal symmetry
and is probably harder to implement in solid state systems.  
However spin-chirality order may 
have important consequences in frustrated magnets\cite{Nagaosa, WenXG}, 
and a realization of spin-chirality 
interactions in cold atom optical lattices has been proposed\cite{Tsomokos}. 

Our model \Equ{equ:Hexplicit} is achieved at second order 
of the perturbation series. Higher order terms become truncation errors
but may be controlled by small parameters 
 $\lambda_{x,y,z}/J_{\rm cluster} \sim \sqrt{|J_{x,y,z}|/J_{\rm cluster}}$.

\section{Conclusions.}\label{sec:conclusion}
We constructed the exactly solvable Kitaev honeycomb model\cite{Kitaev} as 
the exact low energy effective 
Hamiltonian of a spin-1/2 model [equations \Equ{equ:Hexplicit}
 or \Equ{equ:Hexplicit2}] with 
spin-rotation and time reversal symmetry. 
The spin in Kitaev model is represented as the pseudo-spin in the 
two-fold degenerate spin singlet subspace of a cluster of 
four antiferromagnetically coupled spin-1/2 moments. 
The physical spin model is a honeycomb lattice of such four-spin clusters, 
with certain inter-cluster interactions. 
The machinery for the exact mapping to pseudo-spin Hamiltonian was developed 
(see {e.g.} \Tab{table:correspondence}), 
which is quite general and can be used to construct other interesting
(exactly solvable) spin-1/2 models from spin rotation invariant systems.

In this construction the pseudo-spin correlations in the Kitaev model
will be mapped to dimer or spin-chirality correlations in the physical 
spin system. The corresponding picture of the fractionalized Majorana fermion
excitations and Ising vortices still remain to be clarified. 

This exact construction contains high order physical spin interactions, 
which is undesirable for practical implementation. 
We described two possible approaches to reduce this problem: 
generating the high order spin interactions by 
perturbative expansion of the coupling to optical phonon, 
or the magnetic coupling between clusters. 
This perturbative construction will introduce truncation error 
of perturbation series, which may be controlled by 
small expansion parameters. 
Whether these constructions can be experimentally engineered 
is however beyond the scope of this study. 
It is conceivable that other perturbative expansion can also generate 
these high order spin interactions,
but this possibility will be left for future works.

\begin{acknowledgments}
The author thanks Ashvin Vishwanath, Yong-Baek Kim and Arun Paramekanti for 
inspiring discussions, and Todadri Senthil for critical comments. 
The author is supported by the MIT Pappalardo Fellowship in Physics. 
\end{acknowledgments}

\appendix

\section{Coupling between Distortions of a Tetrahedron and the Pseudo-spins}
\label{app:phonon}
In this Appendix we reproduce from Ref.~\cite{TchernyshyovReview} 
the couplings of all tetrahedron distortion modes to the spin system.
And convert them to pseudo-spin notation in the physical spin singlet sector. 

Consider a general small distortion of the tetrahedron, 
the spin Hamiltonian becomes
\be
H_{\rm cluster,\ SL}=
(J_{\rm cluster}/2)(\sum_{\ell}\vecS_{\ell})^2
+J'\sum_{\ell < m} \delta r_{\ell m} (\vecS_{\ell}\cdot\vecS_m)
\label{equ:generalSL}
\ee
where $\delta r_{\ell m}$ is the change of bond length between spins
 $\ell$ and $m$, 
$J'$ is the derivative of $J_{\rm cluster}$ with respect to bond length. 

There are six orthogonal distortion modes of the tetrahedron 
[TABLE~1.1 in Ref.~\cite{TchernyshyovReview}].
One of the modes $A$ is 
 the trivial representation of the tetrahedral group $T_d$;
two $E$ modes form the two dimensional irreducible representation of $T_d$; 
and three $T_2$ modes form the three dimensional irreducible representation.
The $E$ modes are also illustrated in \Fig{fig:QE}. 

The generic couplings in \Equ{equ:generalSL} [second term] 
can be converted to couplings to these orthogonal modes,
\bes
J'(Q^A f^A+Q^E_1 f^E_1+Q^E_2 f^E_2
+Q^{T_2}_1 f^{T_2}_1+Q^{T_2}_2 f^{T_2}_2+Q^{T_2}_3 f^{T_2}_3)
\ees
where $Q$ are generalized coordinates of the corresponding modes,
functions $f$ can be read off from TABLE~1.2 of Ref.~\cite{TchernyshyovReview}.
For the $A$ mode, $\delta r_{\ell m}=\sqrt{2/3}Q^A$, so $f^A$ is
\bes
\begin{split}
f^A=
\ &
\sqrt{2/3}\,
(
\vecS_1\cdot\vecS_2+\vecS_3\cdot\vecS_4+
\vecS_1\cdot\vecS_3
\\ \ & \phantom{\sqrt{2/3}(}
+\vecS_2\cdot\vecS_4+
\vecS_1\cdot\vecS_4+\vecS_2\cdot\vecS_3
).
\end{split}
\ees
The functions $f^E_{1,2}$ for the $E$ modes have been given before
but are reproduced here,
\bes
\begin{split}
f^E_2\ & = (1/2)(
 \vecS_2\cdot\vecS_4+\vecS_1\cdot\vecS_3
-\vecS_1\cdot\vecS_4-\vecS_2\cdot\vecS_3
),
\\
f^E_1\ & = \sqrt{1/12}(
 \vecS_1\cdot\vecS_4+\vecS_2\cdot\vecS_3+\vecS_2\cdot\vecS_4
+\vecS_1\cdot\vecS_3
\\ & \phantom{=\sqrt{1/12}(}
-2\vecS_1\cdot\vecS_2-2\vecS_3\cdot\vecS_4
).
\end{split}
\ees
The functions $f^{T_2}_{1,2,3}$ for the $T_2$ modes are
\bes
\begin{split}
f^{T_2}_1\ & = (\vecS_2\cdot\vecS_3-\vecS_1\cdot\vecS_4),\\
f^{T_2}_2\ & = (\vecS_1\cdot\vecS_3-\vecS_2\cdot\vecS_4),\\
f^{T_2}_3\ & = (\vecS_1\cdot\vecS_2-\vecS_3\cdot\vecS_4)
\end{split}
\ees

Now we can use \Tab{table:correspondence} to convert the 
above couplings into pseudo-spin. 
It is easy to see that 
$f^A$ and $f^{T_2}_{1,2,3}$ are all zero when converted 
to pseudo-spins, namely projected to the physical spin singlet sector. 
But
 $f^E_1=(P_{14}+P_{23}+P_{24}+P_{13}-2P_{12}-2P_{34})/(4\sqrt{3})=-(\sqrt{3}/2)\tau^x$
and 
 $f^E_2=(P_{24}+P_{13}-P_{14}-P_{23})/4=(\sqrt{3}/2)\tau^y$. 
This has already been noted by Tchernyshyov {\etal}\cite{Tchernyshyov},  
only the $E$ modes can lift the degeneracy of the physical spin singlet ground
states of the tetrahedron. 
Therefore the general spin lattice coupling is 
the form of \Equ{equ:HSLcluster} given in the main text.

\section{Derivation of the Terms Generated by Second Order Perturbation 
of Inter-cluster Magnetic Interactions}
\label{app:derive}
In this Appendix we derive the second order perturbations of 
inter-cluster Heisenberg and spin-chirality interactions. 
The results can then be used to construct \Equ{equ:Hmagnetic}. 

First consider the perturbation
 $\lambda\, H_{\rm perturbation}=
 \lambda[\vecS_{j1}\cdot\vecS_{k1}+ r(\vecS_{j2}\cdot\vecS_{k2})]$, 
where $r$ is a real number to be tuned later. 
Due to the fact mentioned in \SSec{ssec:magnetic}, 
the action of $H_{\rm perturbation}$ on 
any cluster singlet state will produce a state 
with total spin-1 for both cluster $j$ and $k$.
Thus the first order perturbation in 
\Equ{equ:PerturbationSeries} vanishes. 
And the second order perturbation term can be greatly simplified: 
operator 
$(1-\mathcal{P}_{jk})
[0-H_{{\rm cluster}\ j}-H_{{\rm cluster}\ k}]^{-1} 
(1-\mathcal{P}_{jk})$ can be replaced by a $c$-number
$(-2 J_{\rm cluster})^{-1}$.
Therefore the perturbation up to second order is
\bes
-\frac{\lambda^2}{2 J_{\rm cluster}}\,
\mathcal{P}_{jk}(H_{\rm perturbation})^2\mathcal{P}_{jk}
\ees
This is true for other perturbations considered later in this Appendix. 
The cluster $j$ and cluster $k$ parts can be separated,
this term then becomes ($a,b=x,y,z$),
\bes
\begin{split}
& 
-\frac{\lambda^2}{2 J_{\rm cluster}}\,\sum_{a,b}
\big [
\mathcal{P}_{j} 
S^a_{j1} S^b_{j1}
\mathcal{P}_{j}
\cdot
\mathcal{P}_{k}
S^a_{k1} S^b_{k1}
\mathcal{P}_{k}
\\ &\quad\quad
+
2 r\,
\mathcal{P}_{j} 
S^a_{j1} S^b_{j2}
\mathcal{P}_{j}
\cdot
\mathcal{P}_{k}
S^a_{k1} S^b_{k2}
\mathcal{P}_{k}
\\ &\quad\quad
+r^2\,
\mathcal{P}_{j}
S^a_{j2} S^b_{j2}
\mathcal{P}_{j}
\cdot
\mathcal{P}_{k}
S^a_{k2} S^b_{k2}
\mathcal{P}_{k}
\big ]
\end{split}
\ees
Then use the fact that 
 $\mathcal{P}_{j}S^a_{j\ell} S^b_{jm}\mathcal{P}_{j}= 
\delta_{ab}
 (1/3)\mathcal{P}_{j}
 (\vecS_{j\ell}\cdot\vecS_{jm})\mathcal{P}_{j}$ by 
spin rotation symmetry, 
the perturbation becomes
\bes
\begin{split}
&
-\frac{\lambda^2}{6 J_{\rm cluster}}
\Big [
\frac{9+9r^2}{16}
+2r\, \mathcal{P}_{jk}(\vecS_{j1}\cdot\vecS_{j2})
(\vecS_{k1}\cdot\vecS_{k2})\mathcal{P}_{jk}
\Big ]
\\
=
\ &
-\frac{\lambda^2}{6 J_{\rm cluster}}
\Big [
\frac{9+9r^2}{16}
+(r/2)\tau^x_j\tau^x_k-r/2
\\ & \phantom{ -\frac{\lambda^2}{6 J_{\rm cluster}} \Big [}
-r\,
\mathcal{P}_{jk}
(\vecS_{j1}\cdot\vecS_{j2}+\vecS_{k1}\cdot\vecS_{k2})
\mathcal{P}_{jk}
\Big ].
\end{split}
\ees
So we can choose
$-(r\,\lambda^2)/(12 J_{\rm cluster})=-J_x$, 
and include the last intra-cluster
 $\vecS_{j1}\cdot\vecS_{j2}+\vecS_{k1}\cdot\vecS_{k2}$ term
in the first order perturbation. 

The perturbation on $x$-links is then (not unique),
\bes
\begin{split}
\lambda_x\, H_{{\rm perturbation},\ x}
=
&
\lambda_x
[\vecS_{j1}\cdot\vecS_{k1}+ \sgn(J_x)\cdot(\vecS_{j2}\cdot\vecS_{k2})]
\\ &
-J_x(\vecS_{j1}\cdot\vecS_{j2}+\vecS_{k1}\cdot\vecS_{k2})
\end{split}
\ees
with $\lambda_x=\sqrt{12 |J_x|\cdot J_{\rm cluster}}$, and 
$r={\rm sgn}(J_x)$ is the sign of $J_x$. The non-trivial terms produced by 
up to second order perturbation will be the $\tau^x_{j}\tau^x_{k}$ term. 
Note that the last term in the above equation commutes with cluster 
Hamiltonians so it does not produce second or higher order perturbations.

Similarly considering the following perturbation on $y$-links, 
$\lambda\,H_{\rm perturbation}=\lambda[\vecS_{j1}\cdot (\vecS_{k3}-\vecS_{k4})
+r\,\vecS_{k1} \cdot (\vecS_{j3}-\vecS_{j4})]$. 
Following similar procedures we get the second order perturbation
from this term
\bes
\begin{split}
& 
-\frac{\lambda^2}{6 J_{\rm cluster}}
\Big [
\frac{9+9r^2}{8}
\\ &\quad
+2r\,
\mathcal{P}_{jk}
[\vecS_{j1}\cdot(\vecS_{j3}-\vecS_{j4})]
[\vecS_{k1}\cdot(\vecS_{k3}-\vecS_{k4})]
\mathcal{P}_{jk}
\\ &\quad
-(3/2)\, 
\mathcal{P}_{jk}
(\vecS_{k3}\cdot\vecS_{k4}
+r^2\,\vecS_{j3}\cdot\vecS_{j4})
\mathcal{P}_{jk}
\Big ]
\\
= \ &
-\frac{\lambda^2}{6 J_{\rm cluster}}
\Big [
\frac{9+9r^2}{8}
+2r\,(3/4)\tau^y_j\tau^y_k
\\ &\quad
-(3/2)\, 
\mathcal{P}_{jk}
(\vecS_{k3}\cdot\vecS_{k4}
+r^2\,\vecS_{j3}\cdot\vecS_{j4})
\mathcal{P}_{jk}
\Big ]
\end{split}
\ees
So we can choose
$-(r\,\lambda^2)/(4 J_{\rm cluster})=-J_y$, 
and include the last intra-cluster
 $\vecS_{k3}\cdot\vecS_{k4}+r^2\,\vecS_{j3}\cdot\vecS_{j4}$ term
in the first order perturbation. 

Therefore we can choose the following perturbation on $y$-links (not unique),
\bes
\begin{split}
&
\lambda_y\, H_{{\rm perturbation},\ y}
\\
=
& 
\lambda_y
[\vecS_{j1}\cdot\vecS_{k1}+ {\rm sgn}(J_y)\cdot
(\vecS_{j3}-\vecS_{j4})\cdot(\vecS_{k3}-\vecS_{k4})]
\\ & 
-|J_y| (\vecS_{j3}\cdot \vecS_{j4}+\vecS_{k3}\cdot \vecS_{k4})
\end{split}
\ees
with $\lambda_y=\sqrt{4 |J_y|\cdot J_{\rm cluster}}$, 
$r=\sgn(J_y)$ is the sign of $J_y$.

The $\tau^z_j\tau^z_k$ term is again more difficult to get. 
We use the representation of $\tau^z$ by spin-chirality \Equ{equ:tauz2}. 
And consider the following perturbation
\bes
H_{\rm perturbation}=\vecS_{j2}\cdot(\vecS_{j3}\times\vecS_{j4})
+r\,\vecS_{k2}\cdot(\vecS_{j3}\times\vecS_{j4})
\ees
The first order term in \Equ{equ:PerturbationSeries} vanishes due to 
the same reason as before. There are four terms in 
the second order perturbation. 
The first one is 
\bes
\begin{split}
& 
\lambda^2\,
\mathcal{P}_{jk}\vecS_{j2}\cdot(\vecS_{k3}\times\vecS_{k4})(1-\mathcal{P}_{jk})
\\ & \times
[0-H_{{\rm cluster}\ j}-H_{{\rm cluster}\ k}]^{-1}
\\ & \times
(1-\mathcal{P}_{jk})\vecS_{j2}\cdot(\vecS_{k3}\times\vecS_{k4})\mathcal{P}_{jk}
\end{split}
\ees
For the cluster $j$ part we can use the same arguments as before, 
the $H_{{\rm cluster}\ j}$ can be replaced by 
a $c$-number $J_{\rm cluster}$.
For the cluster $k$ part, 
consider the fact that $\vecS_{k3}\times\vecS_{k4}$ 
equals to the commutator $-\im[\vecS_{k4},\vecS_{k3}\cdot\vecS_{k4}]$, 
the action of $\vecS_{k3}\times\vecS_{k4}$ on 
physical singlet states of $k$ will also only produce spin-1 state. 
So we can replace the $H_{{\rm cluster}\ k}$ in the denominator 
by a $c$-number $J_{\rm cluster}$ as well. 
Use spin rotation symmetry to separate the $j$ and $k$ parts, 
this term simplifies to
\bes
-\frac{\lambda^2}{6 J_{\rm cluster}} 
\mathcal{P}_{j}\vecS_{j2}\cdot\vecS_{j2}\mathcal{P}_{j}\cdot
\mathcal{P}_{k}(\vecS_{k3}\times\vecS_{k4})\cdot 
(\vecS_{k3}\times\vecS_{k4})\mathcal{P}_{k}.
\ees

Use 
 $(\vecS)^2=3/4$ and
\bes
\begin{split}
&
 (\vecS_{k3}\times\vecS_{k4})\cdot 
 (\vecS_{k3}\times\vecS_{k4})
\\
= \ 
&
\sum_{a,b}(S_{k3}^a S_{k4}^b S_{k3}^a S_{k4}^b
-S_{k3}^a S_{k4}^b S_{k3}^b S_{k4}^a)
\\
= \ 
&
(\vecS_{k3}\cdot \vecS_{k3})(\vecS_{k4}\cdot \vecS_{k4})
-\sum_{a,b}S_{k3}^a S_{k3}^b [\delta_{ab}/2-S_{k4}^a S_{k4}^b ]
\\
= \ 
&
9/16
+(\vecS_{k3}\cdot \vecS_{k4})(\vecS_{k3}\cdot \vecS_{k4})
-(3/8)
\end{split}
\ees
this term becomes
\bes
\begin{split}
&
-\frac{\lambda^2}{6 J_{\rm cluster}}\cdot (3/4)
[3/16+(\tau^x/2-1/4)^2]
\\ \ 
=
&
-(\lambda^2)/(32 J_{\rm cluster})\cdot (2-\tau^x_k).
\end{split}
\ees

Another second order perturbation term $
r^2\lambda^2\,
\mathcal{P}_{jk}\vecS_{k2}\cdot(\vecS_{j3}\times\vecS_{j4})(1-\mathcal{P}_{jk})
[0-H_{{\rm cluster}\ j}-H_{{\rm cluster}\ k}]^{-1}
(1-\mathcal{P}_{jk})\vecS_{k2}\cdot(\vecS_{j3}\times\vecS_{j4})\mathcal{P}_{jk}
$
can be computed in the similar way 
and gives the result
$
-(r^2\,\lambda^2)/(32 J_{\rm cluster})\cdot (2-\tau^x_j)
$.

For one of the cross term
\bes
\begin{split}
& 
r\,\lambda^2\,
\mathcal{P}_{jk}\vecS_{j2}\cdot(\vecS_{k3}\times\vecS_{k4})(1-\mathcal{P}_{jk})
\\ & \times
[0-H_{{\rm cluster}\ j}-H_{{\rm cluster}\ k}]^{-1}
\\ & \times 
(1-\mathcal{P}_{jk})\vecS_{k2}\cdot(\vecS_{j3}\times\vecS_{j4})\mathcal{P}_{jk}
\end{split}
\ees
We can use the previous argument for both cluster $j$ and $k$, 
so $(1-\mathcal{P}_{AB})
[0-H_{{\rm cluster}\ j}-H_{{\rm cluster}\ k}]^{-1}
(1-\mathcal{P}_{jk})$ can be replace by $c$-number $(-2J_{\rm cluster})^{-1}$. 
This term becomes
\bes
-\frac{r\,\lambda^2}{2 J_{\rm cluster}}
\mathcal{P}_{jk}[\vecS_{j2}\cdot(\vecS_{k3}\times\vecS_{k4})]
[\vecS_{k2}\cdot(\vecS_{j3}\times\vecS_{j3})]\mathcal{P}_{jk}.
\ees
Spin rotation symmetry again helps to separate the terms 
for cluster $j$ and $k$, 
and we get 
$
-(r\,\lambda^2)/(32 J_{\rm cluster})\cdot \tau^z_j\tau^z_k
$.

The other cross term 
$r\,\lambda^2\,
\mathcal{P}_{jk}\vecS_{k2}\cdot(\vecS_{j3}\times\vecS_{j4})(1-\mathcal{P}_{jk})
[0-H_{{\rm cluster}\ j}-H_{{\rm cluster}\ k}]^{-1}
(1-\mathcal{P}_{jk})\vecS_{j2}\cdot(\vecS_{k3}\times\vecS_{k4})\mathcal{P}_{jk}
$ gives the same result. 

In summary the second order perturbation from 
 $\lambda [\vecS_{j2}\cdot(\vecS_{j3}\times\vecS_{j4})
+r\,\vecS_{k2}\cdot(\vecS_{j3}\times\vecS_{j4})]$ is
\bes
-\frac{r\,\lambda^2}{16 J_{\rm cluster}}\cdot \tau^z_j\tau^z_k
+\frac{\lambda^2}{32 J_{\rm cluster}}(\tau^x_k+r^2\,\tau^x_j-2r^2-2).
\ees

Using this result we can choose the following perturbation on $z$-links,
\bes
\begin{split}
& 
\lambda_z\, H_{{\rm perturbation},\ z}
\\
=
& 
\lambda_z
[\vecS_{j2}\cdot(\vecS_{k3}\times\vecS_{k4})
+\sgn(J_z)\cdot\vecS_{k2}\cdot(\vecS_{j3}\times\vecS_{j4})]
\\ & 
-|J_z| (\vecS_{j3}\cdot \vecS_{j4}+\vecS_{k3}\cdot \vecS_{k4})
\end{split}
\ees
with $\lambda_z=4\sqrt{|J_z|J_{\rm cluster}}$, 
$r=\sgn(J_z)$ is the sign of $J_z$.  
The last term on the right-hand-side is to cancel 
the non-trivial terms $(r^2\,\tau^x_j+\tau^x_k)\lambda_z^2/(32 J_{\rm cluster})$ 
from the second order perturbation of 
the first term. Up to second order perturbation 
this will produce $-J_z\tau^z_j\tau^z_k$ interactions. 

Finally we have been able to reduce the high order interactions to at most 
three spin terms, the Hamiltonian $H_{\rm magnetic}$ is 
\bes
\begin{split}
H_{\rm magnetic}=
&
\sum_{j}H_{{\rm cluster}\ j}
+ \sum_{x-{\rm links}\ <jk>} \lambda_x H_{{\rm perturbation}\ x}
\\ &
+ \sum_{y-{\rm links}\ <jk>} \lambda_y H_{{\rm perturbation}\ y}
\\ &
+ \sum_{z-{\rm links}\ <jk>} \lambda_z H_{{\rm perturbation}\ z}
\end{split}
\ees
where $H_{{\rm cluster}\ j}$ are given by \Equ{equ:Hcluster}, 
$\lambda_{x,y,z}\,H_{{\rm perturbation}\ x,y,z}$ are given above. 
Plug in relevant equations we get \Equ{equ:Hmagnetic} in \SSec{ssec:magnetic}.

\end{document}